\documentstyle[amsfonts,epsfig,12pt]{article}

\newcommand {\eqref} [1] {(\ref {#1})}

\newcommand {\slsh} [1] {\not{\hbox{\kern-2pt${#1}$}}}

\def\drawbox#1#2{\hrule height#2pt
         \hbox{\vrule width#2pt height#1pt \kern#1pt
               \vrule width#2pt}
               \hrule height#2pt}

\def\Asym#1#2{\vcenter{\vbox{\drawbox{#1}{#2}
               \kern-#2pt       % line up boxes
               \drawbox{#1}{#2}}}}

\newcommand {\beq} {\begin{equation}}
\newcommand {\eeq} {\end{equation}}
  \newcommand {\ber}{\begin{eqnarray*}}
  \newcommand {\eer} {\end{eqnarray*}}
\newcommand {\bea}{\begin{eqnarray}}
  \newcommand {\eea} {\end{eqnarray}}

\newcommand{\Dslash}{\,{\raise.15ex\hbox{/}\mkern-12mu D}}

\begin{document}

%%%%%%%%%%%%%%%%%%%%%%%%%%%%%%%%%%%%%%%%%%%%%%%%%%%%%%%%%%%%%%%%%%%%%%%%%%%%%%%

\begin{titlepage}

\begin{center}
\vspace{1in}
\large{\bf Comments on Mesonic Correlators}\\
\large{\bf in the Worldline Formalism}\\
\vspace{0.4in}
\large{Adi Armoni and Oded Mintakevich}\\
\small{\texttt{a.armoni@swan.ac.uk,\,\,odedm@post.tau.ac.il }}\\
\vspace{0.2in}
\large{{\emph Department of Physics, Swansea University}\\ 
\emph{Singleton Park, Swansea, SA2 8PP, UK}\\}
\vspace{0.3in}
\end{center}

\abstract{We elaborate on how to incorporate mesonic correlators into the worldline formalism. We consider possible applications to QCD-like theories in various dimensions.
 We focus on large-$N_c$ two dimensional QCD (the 't Hooft model) and relate it to a single harmonic oscillator. We also discuss the dependence of the Peskin S-parameter on the number of massless flavors and their representation and compare our expression to the corresponding expression obtained at weak coupling.
 Finally, we use the worldline formalism to discuss how the Veneziano limit of QCD is realized in holography in the limit of small $N_f/N_c$. }

\end{titlepage}

%%%%%%%%%%%%%%%%%%%%%%%%%%%%%%%%%%%%%%%%%%%%%%%%%%%%%%%%%%%%%%%%%%%%%%%%%%%%%%%%
\section{Introduction}

The worldline formalism \cite{Strassler:1992zr} enables to express the QCD fermionic determinant in terms of Wilson
loops. Since in holography Wilson loops are realized by minimal surfaces, the worldline formalism can be
used to incorporate dynamical matter in an arbitrary representation into the gauge/gravity duality framework \cite{Armoni:2008jy}.
Other recent applications of the worldline formalism include the proof of planar equivalence \cite{Armoni:2004ub} and 
an estimate of the size of the conformal window \cite{Armoni:2009jn}.

In this note we wish to discuss how two-point functions of mesonic operators are included in the worldline
formalism. It had been previously discussed in \cite{Vyas:2005wt}. Our main result is that mesonic two-point functions
of the form $\langle \bar \Psi \gamma ^\mu \Psi (x) \, , \,  \bar \Psi \gamma ^\nu \Psi (y) \rangle $
 are computed
by summing all possible (super-)Wilson loops that pass via the points $x$ and $y$.
The purpose of this paper is to discuss various applications of the above result.

The content of the paper is as follows:
in section 2 we briefly review the worldline formalism and study how to incorporate mesonic $k$-point functions in it. We then consider various applications (which are almost unrelated to each other) of our main result. In section 3 we focus on two-dimensional QCD. We consider the 't Hooft large-$N_c$ limit with small $N_f/N_c$. In this limit 2d QCD was solved a long time ago by 't Hooft. It consists of a single Regge trajectory of mesons whose masses are given by $M_n ^2 = \pi g^2 N_c n$. We show that the worldline formalism maps 2d QCD to a single harmonic oscillator and comment on the relation with non-critical string theory. In section 4 we discuss the behavior of the mesonic 2-point function in four dimensional theories at strong coupling. We use our result to calculate the dependence of the Peskin S-parameter on $N_f$ and the matter representation. We compare the expression of the S-parameter at strong coupling with the corresponding weak coupling expression. In section 5 we discuss the Veneziano limit of QCD. We use the worldline formalism to incorporate $k$-point mesonic functions in a holographic description of QCD. We find that the Feynman graphs of QCD in the Veneziano limit resemble at strong coupling string worldsheets, except that the string worldsheet acquire another dimension that corresponds to the holographic RG coordinate. 

\section{Mesonic correlators}

Let us briefly review the worldline formalism \cite{Strassler:1992zr}. The basic idea is to express the fermionic determinant in terms of Wilson loops. The precise relation between the fermionic determinant and Wilson loops is as follows
\beq
\left ( \det i\slsh \!D  \right )^{N_f}  = \exp N_f \Gamma [A] \, ,
\eeq
and 
\beq
Z = \langle \exp N_f \Gamma [A]\rangle_{YM} \, \label{exponent} ,
\eeq

where
\bea
\label{wlineint}
 \Gamma [A] &=&
-{1\over 2} \int _0 ^\infty {dT \over T}
\nonumber\\[3mm]
 &\times&
\int {\cal D} x {\cal D}\psi
\, \exp
\left\{ -\int _{\epsilon} ^T d\tau \, \left ( {1\over 2} \dot x ^\mu \dot x ^\mu + {1\over
2} \psi ^\mu \dot \psi ^\mu \right )\right\}
\nonumber \\[3mm]
 &\times &  {\rm Tr }\,
{\cal P}\exp \left\{   i\int _0 ^T d\tau
\,  \left (A_\mu \dot x^\mu -\frac{1}{2} \psi ^\mu F_{\mu \nu}  \psi ^\nu
\right ) \right\}  \, ,
\eea
with $x^\mu (0)=x^\mu (T)$.
Thus $\Gamma [A]$ is a sum over (super)-Wilson loops. The sum is over contours of all sizes and shapes.

Consider now a two point function of the form
 $\langle \bar \Psi \gamma ^\mu \Psi (x) \, , \,  \bar \Psi \gamma ^\nu \Psi (y) \rangle $. For simplicity we focus on vector meson. The two point function is given by
\beq
 \langle \bar \Psi \gamma ^\mu \Psi (x) \, , \,  \bar \Psi \gamma ^\nu \Psi (y) \rangle = \int {d^d k \over (2\pi)^d} e ^{ ik(x-y)} \left ( k^2 g^{\mu \nu} - k^\mu k^\nu  \right )
 G(k^2) \,.
\eeq
At large-$N$ $G(k^2)$ corresponds to a sum over the vector meson propagators
\beq
G(k^2) = \sum _ n {f_n ^2 \over k^2 - M^2_n}
\eeq

\begin{figure}[!ht]
\centerline{\includegraphics[width=5cm]{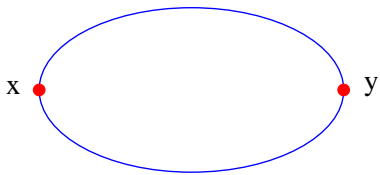}}
\caption{\footnotesize A typical contribution to the mesonic two point function.} \label{w1}
\end{figure}

Let us see how mesonic correlators enter the worldline formalism. Let us couple the current  $\bar \Psi \gamma ^\mu \Psi (x)$ to an external gauge field $B_\mu$, $S=\int d^d x \, B_{\mu} \bar \Psi \gamma ^\mu \Psi$. The two point function is given by
 \beq
\langle \bar \Psi \gamma ^\mu \Psi (x) \, , \,  \bar \Psi \gamma ^\nu \Psi (y) \rangle = {\delta \over \delta B_\mu(x)} {\delta \over \delta B_\nu(y)} Z |_{B=0}
\label{der}
\eeq
Let us expand $\exp N_f \Gamma$ in powers of $N_f$. The first contribution to the above expression \eqref{der} arises from the ${\cal O}(N_f)$ term.  
The derivatives with respect to $B$ project out Wilson loops that do not pass via the points $x$ and $y$. Hence to  ${\cal O}(N_f)$ we obtain
\beq
{\delta \over \delta B_\mu(x)} {\delta \over \delta B_\nu(y)} Z |_{B=0}=\int DA^\mu (\exp -S_{\rm YM}) \times N_f \Gamma_{x,y}[A] \, ,
\eeq
where $\Gamma_{x,y}[A]$ includes all Wilson loops that pass via $x$ and $y$ as depicted in figure \eqref{w1}.

Let us restrict ourselves to the bosonic part of $\Gamma$. Omitting the worldline fermions is justified if the spin of the quarks in neglected. We obtain
\bea
  \int {d^2 k \over (2\pi)^2} G(k^2) (\exp ik(x-y)) = \int {\cal D} A^\mu (\exp -S_{\rm YM})\times \nonumber \\
 {N_f\over 2} \int {dT \over T} 
\oint_{x,y} {\cal D} x 
\, (\exp -\int \, {1\over 2} \dot x _\mu ^2 ) {\rm Tr }\, (\exp i\int
\, A_\mu \dot x^\mu )   \, \label{bosonic}.
\eea

In the following sections we will discuss various applications of the expression \eqref{bosonic}.

\begin{figure}[!ht]
\centerline{\includegraphics[width=5cm]{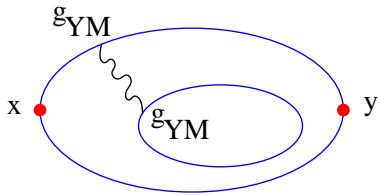}}
\caption{\footnotesize A second order contribution to a mesonic two-point function. The wavy line represents a gluon exchange between the Wilson loops.} \label{w2}
\end{figure}

Let us elaborate on higher order terms in the expansion of \eqref{exponent}. Naively the expansion is in powers of $N_f$, but it is actually an expansion in powers of $N_f/N_c$. The reason, as explained in \cite{Armoni:2009jn}, is that we need to evaluate connected Wilson loops of the form $\langle W_1 W_2 ... W_l \rangle _{\rm conn.}$. Going to a higher order in the expansion involves a higher power of $N_f$, but since we need to connect a Wilson loop to the rest of the diagram at least one gluon exchange is required. The gluon coupling to two Wilson loops involves $g^2_{\rm YM}$, or $\lambda /N_c$. In order to clarify this issue in a simple case, a typical ${\cal O} \left ( (N_f /N_c)^2 \right )$ diagram is depicted in figure \eqref{w2}.
 
\section{Two dimensional QCD}

Two dimensional QCD in the large-$N_c$ limit was solved a long time ago in the seminal work of 't Hooft \cite{'tHooft:1974hx}. The solution of the model consists of a single Regge trajectory of mesons with an asymptotic (large radial excitation) spectrum of the form $M^2_n = \pi g^2 N_c n$. 

Consider now the expression \eqref{bosonic}. It is the leading ${\cal O}(N_f/N_c)$ contribution. In the 't Hooft limit of large-$N_c$ and fixed $N_f$ it is valid to neglect higher order corrections, hence \eqref{bosonic} is the large-$N_c$ leading bosonic truncation of the partition function.

Let us carry out the integration over the gauge field. In two dimensions the expectation value of a Wilson loop, of any shape and size, respects an area law
\beq
\langle {\rm Tr }\, (\exp i\int \, A_\mu  x^\mu ) \rangle = \exp -\sigma A \, ,
\eeq 
where the string tension is given by $\sigma = g^2 N_c$. Hence the bosonic part of the worldline action takes the form
\beq
 {N_f\over 2} \int {dT \over T} \int {\cal D} x 
\, \exp (-\int _{\partial {\cal A}} \, d\tau {1\over 2} \dot x _\mu ^2  -\sigma \int _{\cal A}
\, dx dy) \, .
\eeq
By using Green's theorem we can convert the surface integral into a contour integral
\beq
\int _{\cal A} dx dy = {1\over 2} \int _{\partial \cal A} xdy - ydx \equiv {1\over 2} \int _{\partial \cal A} d\tau \epsilon _{\mu \nu} x^\mu \dot x^\nu
\eeq
The partition function hence takes the form
 \beq
 {N_f\over 2} \int {dT \over T} \int {\cal D} x 
\, \exp -\int \, d\tau ({1\over 2} \dot x _\mu ^2  - {1\over 2} \sigma \epsilon _{\mu \nu} x^\mu \dot x^\nu )\, . \label{landau}
\eeq
Interestingly the above Lagrangian \eqref{landau} describes the motion of
 an electron in a constant magnetic field of strength $B=\sigma$. This system was analyzed a long time ago by Landau and the corresponding energy levels are those of an harmonic oscillator, with $\omega = B = \sigma$, hence
\beq
E_n = \sigma (n+{1\over 2}) \,.
\eeq
The highly excited energy levels $E_n \rightarrow g^2 N_c n$ should be compared
 with the highly excited masses of the 't Hooft model $M^2 _n \rightarrow \pi g^2 N_c n$. We therefore identify $E_n$ with $M^2_n / \pi$.
 
Few comments are in order. In our mapping of the 't Hooft model to an harmonic oscillator, we have neglected the worldline fermions. Such an approximation is equivalent to neglecting the spin of the two-dimensional fermions, namely to a model where the matter is scalar. Neglecting the spin is valid for highly excited mesons where the contribution to the mass arises from the QCD-string and not from its endpoints.

\begin{figure}[!ht]
\centerline{\includegraphics[width=5cm]{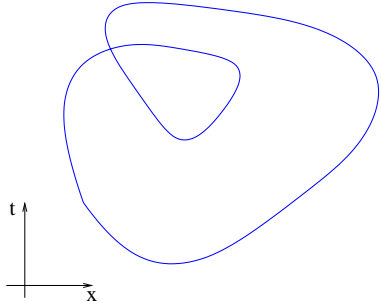}}
\caption{\footnotesize A self intersecting Wilson loop.} \label{intersecting}
\end{figure}

 We have also neglected self intersecting Wilson-loops\footnote{We wish to thank Y. Makeenko for a discussion about self intersecting loop, a comprehensive treatment of
 these loops in this model is found in \cite{Strominger:1980xa}. }. 
 We expect that such loops will be negligible for high excitations. The intuitive explanation is as follows: consider an intersecting loop, as depicted in figure \eqref{intersecting}. The space-time picture is of a string whose worldsheet is singular. While we cannot prove it, we believe that singular worldsheets do not dominate the highly excited masses. On the contrary: the partition function is expected to be dominated by smooth string worldsheets. In fact, a holographic calculation of the QCD string tension does not involve intersecting Wilson loops. Thus, holography supports our intuitive explanation that while low-lying states could be affected by intersecting loops, the contribution to the asymptotic spectrum from those loops is negligible.

Note that the spectrum of 2d QCD consists of a single Regge trajectory. There is no asymptotic Hagedorn density of states in 2d. This is in agreement with the fact that the spectrum of a one-dimensional harmonic oscillator is not degenerate. 

It would be interesting to generalize our result to 4d. There are few difficulties in doing so. The first is that not all the Wilson loops that contribute to the partition function enclose a large area, hence an area law is not justified in this case. Another problem is that a generic Wilson loop does not rest on a plane. It is nevertheless tempting to guess that it is possible to generalize our 
result by using {\it infinitely} many harmonic oscillators - namely by a string.

The last comment is that there have been attempts to describe the 't Hooft model by a dual string \cite{Narayanan:2005gh,Katz:2007br}. The precise relation with string theory is not known yet, but it is reasonable to assume that the dual model should be a 3d non-critical string theory with an IR cut-off along the 'radial' direction.
Our finding that the 't Hooft model can be described (at least the asymptotic mass levels) by an harmonic oscillator, suggests that perhaps one should start by looking for the string dual of a single 1d harmonic oscillator.

\section{S-Parameter}

A well known problem that plague models of Technicolor is that the electro-weak observable receive oblique corrections from the 
hidden Technicolor sector. 
In their paper \cite{Peskin}, Peskin and Takauchi showed how these corrections can be conveniently summed up into three parameters  S, T and U 
which are predictions of each specific model.  These parameters spans a 3d space constrained by  the experimental value of the electro-weak
observables. 
Among these three parameters the S parameter stands out in its importance not only because it is the most reliably estimated,
but also because T and U arise  from interactions that breaks the $SU(2)$ custodial symmetry and so are only relevant in extended Technicolor models
and in most cases one can show (perturbatively) that they are suppressed compared to S
\footnote{ Perturbative estimation (such as eq. (\ref{oneloop})) shows that S measure the size of the hidden sector, and by definition it is isospin independent.}.
Because of that the constraints on the value
of S are more accurate and robust then those on T and U
\footnote{Phenomenology extending the Technicolor is crucial, unfortunately it also turn things too complicated for the current technology we posses.}.\\

The S-parameter is defined as 

\beq\label{S_definition}
S=4\pi\frac{\partial}{\partial Q}\int d^4x e^{iqx}[ \Pi_V(x)-\Pi_A(x)]\bigg|_{Q=0}
\eeq
Where $Q=-q^2$ and 
\beq 
\Pi^{\mu\nu}_{ab}(x)=\langle J^{\mu}_a(x)J^{\nu}_b(0)\rangle=(g_{\mu\nu}-\frac{\partial _{\mu}\partial_{\nu}}{\partial^2})\Pi_{ab}(x)
\eeq

In a basic model  based on a gauge group $SU(N)$ with $N_f$ massless techniquarks transforming in a representation R, such as \cite{Dietrich:2005jn},
a one-loop analysis gives \cite{Sannino:2004qp}
\beq
S={1 \over 12\pi} N_f \dim R \, . \label{oneloop}
\eeq
But this result should be considered only as a hint, since the Technicolor sector is tuned to be strongly interacting in the relevant regime  ($\Lambda_{Tc}>\Lambda_{QCD}$).
A variety of non-perturbative approaches had been applied to this problem \cite{Peskin}, and
recently even the gauge/gravity duality was applied to it \cite{holtech}.
In this section we wish to apply the worldline formalism to estimate the S-parameter in the strongly coupled regime. 

According to our analysis a mesonic two-point function can be written as a sum over Wilson loops as in \eqref{bosonic}. Namely,
\beq
\Pi(x)= N_f \sum _{\cal C} \alpha _{\cal C} \langle W_R(x)\rangle \, . \label{Piwilson}
\eeq
where $ \langle W_R(x)\rangle $ is a Wilson loop passing through two point $x$ and $0$.
Inserting this into (\ref{S_definition}) we find
\beq
S = 4\pi N_f \sum _{\cal C} \alpha _{\cal C}\frac{\partial}{\partial Q}\int d^4x e^{iqx} [\langle W^V_R(x) \rangle-\langle W^A_R(x) \rangle] \bigg|_{Q=0}\, . \label{Swilson}
\eeq  
The dependence on the representation enters solely via the Wilson loop representation. Wilson loops expectation values are proportional to $\dim R$. Moreover,
large Wilson loops exhibit an area law with a string tension that depends on the $N$-ality of the representation
\beq
 \langle W_R \rangle = (\dim R) \exp -\sigma _k {\cal A} \,.
\eeq
 $\sigma_k$ is the string tension of a representation $R$ of $N$-ality $k$ in {\it pure} Yang-Mills theory. 
Therefore, under the assumption that the S-parameter is dominated by confining
Wilson loops (Wilson loops whose vev exhibits an area law) we arrive to the conclusion that
\beq
S=C_k N_f \dim R \, , \label{Sresult}
\eeq 
where $C_k$ is an unknown proportionality coefficient that depends on the
$N$-ality of the representation. In particular it is the same coefficient 
for the two-index symmetric, antisymmetric and the adjoint representation of $SU(N)$, but it differs from the coefficient of the fundamental representation. Since the string tension increases with $k$, $C_k \sim \exp -\sigma_k {\cal A}$ decreases with $k$. In particular, $C_2<C_1$. The latter observation might be useful for technicolor theories based on the symmetric representation \cite{Sannino:2004qp}.  

It is interesting to compare our result \eqref{Sresult} with the perturbative (one-loop) result \eqref{oneloop}: both are proportional to $N_f \dim R$, but in our case the proportionality coefficients $C_k$ are unknown and there is no reason to expect them to be $k$ independent. 

Finally we wish to summarize the assumptions that were made in arriving at \eqref{Sresult}. (i). We assumed that the sum is dominated by large Wilson loops. (ii). We considered only the ${\cal O}(N_f)$ contribution. This is valid when the expansion parameter $N_f C(R)/N_c $ is small.

\section{Veneziano limit of QCD and string theory}

In this section we discuss the Veneziano limit of QCD. We use the worldline formalism to comment on the dual description in string theory.

Consider QCD with an $SU(N_c)$ gauge group and $N_f$ flavors of fermions
transforming in the fundamental representation. This theory in the limit
$N_c \rightarrow \infty$, $N_f/N_c$ and $g^2 N_c$ fixed was analyzed along time ago by Veneziano \cite{Veneziano:1976wm}.

\begin{figure}[!ht]
\centerline{\includegraphics[width=8cm]{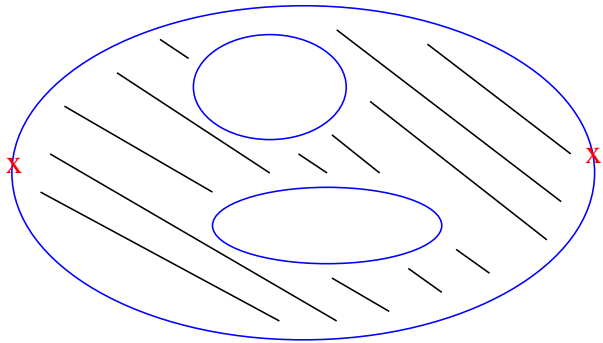}}
\caption{\footnotesize A perturbative planar contribution to the mesonic two point function. Meson insertions are made on the boundary. Fermion loops create windows in the diagram.} \label{te}
\end{figure}

In the above limit (``Veneziano limit'') the dependence of a mesonic $k$-point function on $N_c$ and $N_f$ is \cite{Veneziano:1976wm} (see \cite{Nunez:2010sf} for a recent review)
\beq
\langle M(x_1) M(x_2) ... M(x_k) \rangle \sim \left ({N_f \over N_c} \right )^w
N_c^{(2-{k\over 2}-2h-b)}  \,, \label{kpoint}
\eeq
where $w$ is the number of windows (fermionic loops), $b$ is the number of boundaries where the mesonic operators are inserted and $h$ is the number of holes. A typical diagram is depicted in figure \eqref {te}. Note that the dependence on $N_f$ is solely via the windows in the diagram, created by fermionic loops.

Let us see how the above \eqref{kpoint} dependence is captured by the worldline formalism and holography. 

\begin{figure}[!ht]
\centerline{\includegraphics[width=8cm]{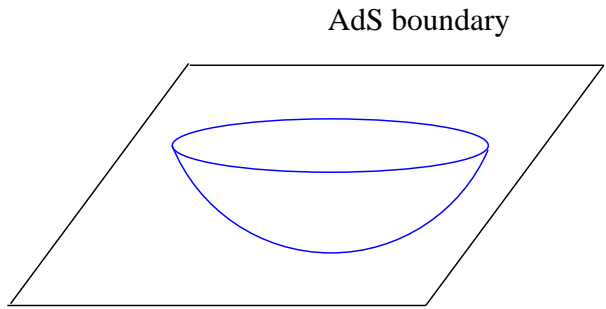}}
\caption{\footnotesize A Wilson loop as a minimal surface in AdS/CFT. Dynamical matter at ${\cal O}(N_f)$ is added to the holographic description by summing over all possible string worldsheets that terminate on boundary. } \label{hol1}
\end{figure}

In the holographic dual Wilson loops are described by string worldsheets that terminate on the boundary of the AdS space (or in general on the boundary of the gravity dual manifold) \cite{Maldacena:1998im}. The worldline formalism tells us that in order to incorporate dynamical matter in the partition function, one has to sum over all sizes and shapes of Wilson loops. Therefore, in order to incorporate dynamical matter in the holographic description, one has to sum over all string worldsheets that terminate on the boundary \cite{Armoni:2008jy}, as depicted in fig. \eqref{hol1}.

In order to compute mesonic $k$-point functions, it is necessary to sum over all Wilson loops that pass via $x_1, x_2, ... , x_k$. The leading holographic contribution is given by a minimal surface that terminates on the AdS (or the dual gravity manifold, in general) boundary and pass via  $x_1, x_2, ... , x_k$. 

In order to compute a contribution at ${\cal O}(N_f^l)$, the worldline prescription is that it is necessary to consider $\langle W_1 W_2 ... W_l \rangle _{\rm conn.}$, namely the connected $l$-point function of Wilson loop operators. The holographic description is via a minimal surface with $l$ loops on the AdS boundary. It is interesting to compare it with the perturbative analysis of Veneziano. The diagrams look almost identical, except that at strong coupling (holographic description) we acquire another dimension and while the holes live on the boundary, the string worldsheet is pulled into the holographic radial direction.

\begin{figure}[!ht]
\centerline{\includegraphics[width=8cm]{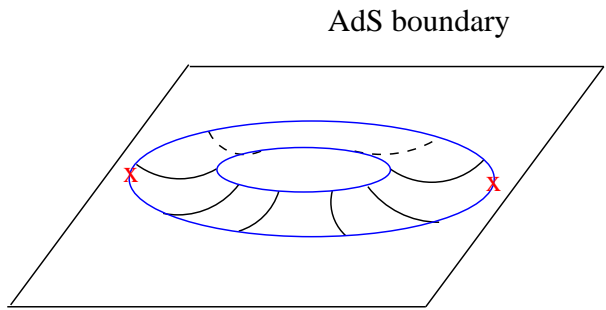}}
\caption{\footnotesize A typical contribution to a mesonic two-point function. The mesonic insertions, denoted by the letter x, sit on one of the Wilson loops. The contribution depicted in the above figure contains one window.} \label{hol2}
\end{figure}

In figure \eqref{hol2} we consider a typical contribution to a mesonic $2$-point
function.

It is easy to see how the dependence of the mesonic $k$-point function on $N_c$ and $N_f$ (eq.\eqref{kpoint}) is correctly captured by the holographic description based on the worldline formalism. The dependence on $h$, $N_c ^{(2-2h)}$ is due to the standard genus expansion in a closed string theory. These are bulk closed string loops. As we already noted each window adds a factor of $N_f/N_c$ (see a detailed discussion in \cite{Armoni:2009jn}), hence the factor $(N_f / N_c)^w$. The dependence $N_c^{(-k/2 -b)}$ is essentially due to the normalization of the mesonic operators.

Thus, we propose that the dual description of QCD in the Veneziano limit is in terms of a closed string theory. The propagating degrees of freedom in the bulk are closed strings only. In addition, one should add string worldsheets that terminate on the boundary. 

We would like to stress that our description makes sense only when $N_f/N_c$ is small and a perturbative expansion around the confining Yang-Mills vacuum is valid. Presumably when $N_f/N_c \sim {\cal O}(1)$ there is no string dual, unless the flavor symmetry is gauged with a large 't Hooft coupling. Such a case was recently discussed in \cite{Gadde:2009dj}. 
 
\section{Summary}

In this short note we derived a prescription, based on the worldline formalism, for a calculation of Mesonic correlators at strong coupling. We focused on three applications: 2d QCD, the Peskin S-parameter and a realization of the Veneziano limit of QCD in holographic duals.

The present work can be continued in various directions. It will be interesting to compute the $k$-point function of mesonic operators in the 't Hooft model by using the worldline formalism and compare it to known results. 
It is important to know whether one can actually use the worldline formalism for holographic computations beyond qualitative estimates. Another important issue is the importance of the contribution of self-intersecting Wilson loops. We postpone these questions for a future work.
 
\vspace{0.5cm}

{\it \bf Acknowledgements.} We wish O. Aharony, M. Field, Y. Frishman, Z. Komargodski, D. Kosower, C. Nunez, M. Piai and J. Sonnenschein for fruitful discussions. A.A. wishes to thank the particle physics group at the Weizmann institute for the kind and warm hospitality, where most of this work has been done. A.A. was a fellow of the Feinberg Foundation Visiting Faculty Program at the Weizmann institute. O.M was supported by Britain-Israel research and academic exchange (BIRAX).

%%%%%%%%%%%%%%%%%%%%%%%%%%%%%%%%%%%%%%%%%%%%%%%%%%%%%%%%%%%%%%%%%%%5


\begin{thebibliography}{99}

%\cite{Strassler:1992zr}
\bibitem{Strassler:1992zr}
  M.~J.~Strassler,
  ``Field theory without Feynman diagrams: One loop effective actions,''
  Nucl.\ Phys.\  B {\bf 385}, 145 (1992)
  [arXiv:hep-ph/9205205].
  %%CITATION = NUPHA,B385,145;%%

%\cite{Armoni:2008jy}
\bibitem{Armoni:2008jy}
  A.~Armoni,
  ``Beyond The Quenched (or Probe Brane) Approximation in Lattice (or Holographic) QCD,''
  Phys.\ Rev.\  D {\bf 78}, 065017 (2008)
  [arXiv:0805.1339 [hep-th]].
  %%CITATION = PHRVA,D78,065017;%%

%\cite{Armoni:2004ub}
\bibitem{Armoni:2004ub}
  A.~Armoni, M.~Shifman and G.~Veneziano,
  ``Refining the proof of planar equivalence,''
  Phys.\ Rev.\  D {\bf 71}, 045015 (2005)
  [arXiv:hep-th/0412203].
  %%CITATION = PHRVA,D71,045015;%%

%\cite{Armoni:2009jn}
\bibitem{Armoni:2009jn}
  A.~Armoni,
  ``The Conformal Window from the Worldline Formalism,''
  Nucl.\ Phys.\  B {\bf 826}, 328 (2010)
  [arXiv:0907.4091 [hep-ph]].
  %%CITATION = NUPHA,B826,328;%%


%\cite{Vyas:2005wt}
\bibitem{Vyas:2005wt}
  V.~Vyas,
  ``Super Wilson loops in planar QCD,''
  arXiv:hep-th/0511176.
  %%CITATION = HEP-TH/0511176;%%

%\cite{'tHooft:1974hx}
\bibitem{'tHooft:1974hx}
  G.~'t Hooft,
  ``A Two-Dimensional Model For Mesons,''
  Nucl.\ Phys.\  B {\bf 75}, 461 (1974).
  %%CITATION = NUPHA,B75,461;%%

%\cite{Strominger:1980xa}
\bibitem{Strominger:1980xa}
  A.~Strominger,
  ``Loop Space Solution Of Two-dimensional Qcd,''
  Phys.\ Lett.\  {\bf B101}, 271 (1981).
  
%\cite{Narayanan:2005gh}
\bibitem{Narayanan:2005gh}
  R.~Narayanan and H.~Neuberger,
  ``The quark mass dependence of the pion mass at infinite N,''
  Phys.\ Lett.\  B {\bf 616}, 76 (2005)
  [arXiv:hep-lat/0503033].
  %%CITATION = PHLTA,B616,76;%%

%\cite{Katz:2007br}
\bibitem{Katz:2007br}
  E.~Katz and T.~Okui,
  ``The 't Hooft Model As A Hologram,''
  JHEP {\bf 0901}, 113 (2009)
  [JHEP {\bf 0901}, 013 (2009)]
  [arXiv:0710.3402 [hep-th]].
  %%CITATION = JHEPA,0901,013;%%

\bibitem{Peskin}
  M.~E.~Peskin and T.~Takeuchi,
``A New constraint on a strongly interacting Higgs sector,''
  Phys.\ Rev.\  Lett. {\bf 65}, 8 (1990)
 % \bibitem{Peskin}
  ; M.~E.~Peskin and T.~Takeuchi,
  ``Estimation of oblique electroweak corrections,''
  Phys.\ Rev.\  D. {\bf 46}, 1 (1992)

%\cite{Dietrich:2005jn}
\bibitem{Dietrich:2005jn}
  D.~D.~Dietrich, F.~Sannino and K.~Tuominen,
  ``Light composite Higgs from higher representations versus electroweak 
precision measurements: Predictions for LHC,''
  Phys.\ Rev.\  D {\bf 72}, 055001 (2005)
  [arXiv:hep-ph/0505059].
  %%CITATION = PHRVA,D72,055001;%%

%\cite{Sannino:2004qp}
\bibitem{Sannino:2004qp}
  F.~Sannino and K.~Tuominen,
  ``Techniorientifold,''
  Phys.\ Rev.\  D {\bf 71}, 051901 (2005)
  [arXiv:hep-ph/0405209].
  %%CITATION = PHRVA,D71,051901;%%


\bibitem{holtech}
  K.~Haba, S.~Matsuzaki and K.~Yamawaki,
  ``$S$ Parameter in the Holographic Walking/Conformal Technicolor,''
  Prog.\ Theor.\ Phys.\  {\bf 120}, 691 (2008)
  [arXiv:0804.3668 [hep-ph]].
  %%CITATION = PTPKA,120,691;%%

%\cite{Hirayama:2007hz}
%\bibitem{Hirayama:2007hz}
  T.~Hirayama and K.~Yoshioka,
  ``Holographic Construction of Technicolor Theory,''
  JHEP {\bf 0710}, 002 (2007)
  [arXiv:0705.3533 [hep-ph]].
  %%CITATION = JHEPA,0710,002;%%

%\cite{Piai:2006hy}
%\bibitem{Piai:2006hy}
  M.~Piai,
  ``Precision electro-weak parameters from AdS(5), localized kinetic terms  and
  anomalous dimensions,''
  arXiv:hep-ph/0608241.
  %%CITATION = HEP-PH/0608241;%%



%\cite{Agashe:2007mc}
%\bibitem{Agashe:2007mc}
  K.~Agashe, C.~Csaki, C.~Grojean and M.~Reece,
  ``The S-parameter in holographic technicolor models,''
  JHEP {\bf 0712}, 003 (2007)
  [arXiv:0704.1821 [hep-ph]].
  %%CITATION = JHEPA,0712,003;%%

%\cite{Carone:2006wj}
%\bibitem{Carone:2006wj}
  C.~D.~Carone, J.~Erlich and J.~A.~Tan,
  ``Holographic Bosonic Technicolor,''
  Phys.\ Rev.\  D {\bf 75}, 075005 (2007)
  [arXiv:hep-ph/0612242].
  %%CITATION = PHRVA,D75,075005;%%

%\cite{Hirn:2006nt}
%\bibitem{Hirn:2006nt}
  J.~Hirn and V.~Sanz,
  ``A negative S parameter from holographic technicolor,''
  Phys.\ Rev.\ Lett.\  {\bf 97}, 121803 (2006)
  [arXiv:hep-ph/0606086].
  %%CITATION = PRLTA,97,121803;%%

%\cite{Hong:2006si}
%\bibitem{Hong:2006si}
  D.~K.~Hong and H.~U.~Yee,
  ``Holographic estimate of oblique corrections for technicolor,''
  Phys.\ Rev.\  D {\bf 74}, 015011 (2006)
  [arXiv:hep-ph/0602177].
  %%CITATION = PHRVA,D74,015011;%%

%\bibitem{Carone:2007md}
  C.~D.~Carone, J.~Erlich and M.~Sher,
  ``Holographic Electroweak Symmetry Breaking from D-branes,''
  Phys.\ Rev.\  D {\bf 76}, 015015 (2007)
  [arXiv:0704.3084 [hep-th]].
  %%CITATION = PHRVA,D76,015015;%%


%\cite{Veneziano:1976wm}
\bibitem{Veneziano:1976wm}
  G.~Veneziano,
  ``Some Aspects Of A Unified Approach To Gauge, Dual And Gribov Theories,''
  Nucl.\ Phys.\  B {\bf 117}, 519 (1976).
  %%CITATION = NUPHA,B117,519;%%

%\cite{Nunez:2010sf}
\bibitem{Nunez:2010sf}
  C.~Nunez, A.~Paredes and A.~V.~Ramallo,
  ``Unquenched flavor in the gauge/gravity correspondence,''
  arXiv:1002.1088 [hep-th].
  %%CITATION = ARXIV:1002.1088;%%

%\cite{Maldacena:1998im}
\bibitem{Maldacena:1998im}
  J.~M.~Maldacena,
  ``Wilson loops in large N field theories,''
  Phys.\ Rev.\ Lett.\  {\bf 80}, 4859 (1998)
  [arXiv:hep-th/9803002].
  %%CITATION = PRLTA,80,4859;%%


%\cite{Gadde:2009dj}
\bibitem{Gadde:2009dj}
  A.~Gadde, E.~Pomoni and L.~Rastelli,
  ``The Veneziano Limit of $N=2$ Superconformal QCD: Towards the String Dual of
$N=2$ $SU(N_c)$ SYM with $N_f =2 N_c$,''
  arXiv:0912.4918 [hep-th].
  %%CITATION = ARXIV:0912.4918;%%


\end{thebibliography}
\end{document}